# Supercooled fog as a natural laboratory for studying ice formation in clouds

*Lea B. Weber and Franz Conen*

Department of Environmental Sciences, University of Basel, Switzerland

*Correspondence to:* Franz Conen (franz.conen@unibas.ch)

## Abstract

The ice phase in clouds contributes largely to uncertainties in global climate models partly due to a lack of atmospheric observations. At moderate supercooling ice nucleating particles (INP) and ice particles (IP) are present in small concentrations and a large volume of air is necessary for observation. Here, we report on initial observations of IP in supercooled fog with a new setup. We use a 0.3 m wide, vertical curtain of light and a camera pointing perpendicularly at it to record light scattered by IP formed in radiation fog near the ground at temperatures between -3 °C and -9 °C. Deposition rates of IP were several times larger than expected from number concentrations of INP found on $PM_{10}$-filters at a nearby air quality monitoring station. The discrepancy might be explained by secondary ice formation through the fragmentation of freezing water droplets, the entrainment of INP from above the fog layer through settling, the loss or deactivation of INP on $PM_{10}$-filters prior to analysis, or radiative cooling of INP below the temperature of the surrounding air. In summary, radiation fog constitutes an easily accessible form of a supercooled cloud, in which observations can be made for long enough to quantify the deposition rate of rare IP produced in a natural environment.

## 1   Introduction

The latest report of the Intergovernmental Panel on Climate Change (IPCC) states that clouds and aerosols are still the most important contributors to uncertainties in estimates and interpretations of the change in the Earth's energy balance (Boucher et al., 2014). Despite considerable progress in the understanding of different ice nucleation processes, the ice phase in clouds still contributes to large uncertainties in the prediction of radiative forcing with climate models (Boose, 2016, III). According to estimates, more than 50% of global precipitation forms in ice containing clouds (Field & Heymsfield, 2015; Mülmenstädt et al., 2015). Ice formation above -35 °C requires ice nucleating particles (INP), aerosols of natural or anthropogenic origin. Depending on temperature range and vapour saturation, mineral particles (e.g. soil dust), soot, marine aerosols, and biological particles such as pollen, plant residues, bacteria, and fungal spores are active as INP (Murray et al., 2012). INP concentrations higher than 1000-10'000 $m^{-3}$ can be found in the atmosphere only below -15 °C where mineral particles are very active as ice nucleators (DeMott & Prenni, 2010). At temperatures above -10 °C only biological INP, whose concentrations have been estimated to be <0.1 to 90 $m^{-3}$ at -10 °C, are active (DeMott & Prenni, 2010). However, the number of IP in clouds frequently exceeds that of INP (Field et al., 2016). The small number concentrations of IP and INP active at moderate supercooling make investigating this issue in clouds extremely challenging.

Fog is a cloud touching the ground and therefore is easily accessible for on-site measurements. In mountain areas atmospheric inversions are common especially in winter. In the closed La Brévine valley in the Swiss Jura mountains for instance, the specific topography leads to a strong sheltering effect, and inversions due to radiative cooling are frequent. At night the valley floor and the air near the ground quickly lose heat while the topography prevents mixing of warmer air from above. As atmospheric stability increases, the developing cold air pool is decoupled from wind flows from outside the valley and vertical mixing is prevented (Vitasse et al., 2016). The strongest inversion observed by Vitasse et al. (2016) in the winter 2014-2015 reached a temperature difference of 6 °C on the lowest 20 m. Under such conditions it is not uncommon for supercooled fog to form near the ground. Such fog constitutes an accessible form of a supercooled cloud, in which observations can be made over a period of hours. Due to the strong inversion the system is relatively closed compared to other low or mid-level clouds.

Data on IP and INP active at moderate supercooling is rare. Therefore, the aim of this study was to test a procedure for detecting small numbers of IP in a large volume of ambient air and to compare its results with measured number concentrations of INP active at the temperature at which the observations were made.

## 2  Methods

### 2.1  Site description

The La Brévine valley is a narrow, closed valley in the Jura Mountains in Switzerland, ranging from 1033 to 1308 m a.s.l (Figure 1). The slopes of the valley are mainly covered by evergreen forests whereas the valley floor is mostly used as meadows for mowing or pasture. The local MeteoSchweiz weather station in La Brévine holds the record of the coldest temperature ever measured in Switzerland: -41.8 °C in January 1987.

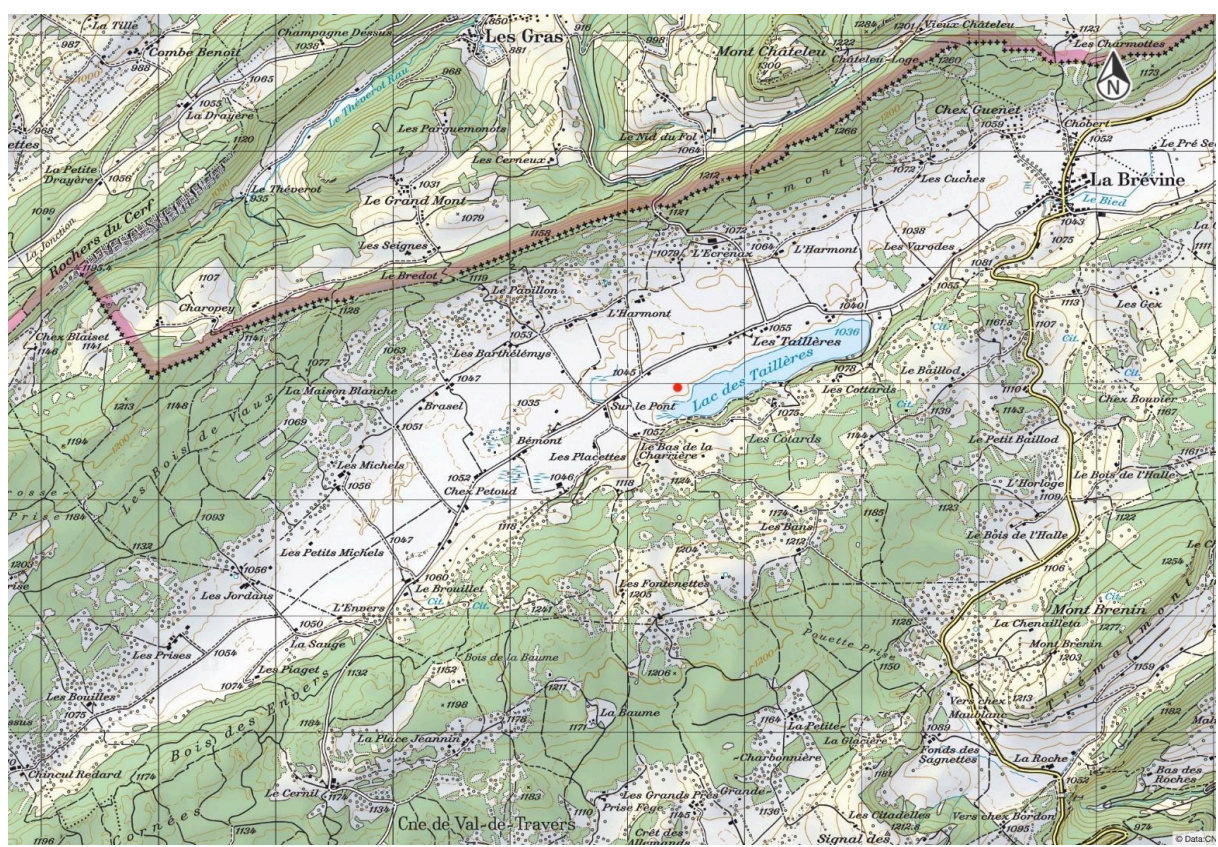

*Figure 1: Measurement location (red dot) in the Vallée de la Brévine (National Map 1:50'000, Swiss Geoportal, WWW). Gridlines are at 1000 m distance.*

Temperatures below -20°C are common in winter and the Vallée de la Brévine is therefore nicknamed "Siberia of Switzerland" (Vitasse et al., 2016). For our measurements, we chose a location on the valley bottom (46°57'55.58"N, 6°33'49.84"E, 1042 m a.s.l.).

## 2.2 Observations of IP and INP

During inversion nights in the winter 2015-2016 we investigated the occurrence of IP in radiation fog with a setup consisting of a white LED light unit for line scan applications pointing vertically into the sky (model CCS LNSP-300SW, light-emitting surface: 300 x 21 mm, 400'000 lx; Stemmer Imaging AG, Pfäffikon, Switzerland) and an industrial camera (model DMK 72BUC02, chip: 1/2.5" Micron CMOS; The Imaging Source, Bremen, Germany) pointing perpendicularly into the homogenous curtain of light (Figure 2). The camera has a resolution of 2592 x 1944 pixels and was equipped with a c-mount lens (model C1614-M 16mm/f1.4, set to 16 / 0.6; PENTAX, Tokyo, Japan). The horizontal distance between camera and light unit was about 80 cm. The light unit was fixed about 1 m above the ground, and the camera was fixed about 20 cm higher than the light-emitting surface. Between camera and light, temperature and relative humidity were measured with a humidity and temperature probe (model HUMICAP HMP155; Vaisala, Helsinki, Finland).

Ice particles forming in the fog become visible through the light they scatter when passing through the light curtain (Yagi, 1970). They leave white traces on an otherwise black image taken by the camera with an exposure time of 2 seconds (20 pictures a minute). The pictures are analysed visually for white traces and IP retraced manually and counted using the GIS programme ArcMap. In order to compare IP and INP abundance the deposition rates of IP are calculated from IP number concentrations as described in Chapter 3. The IP deposition rates are the rates at which IP are deposited to the ground.

Concentrations of INP were determined on $PM_{10}$-filters of the air quality monitoring station Chaumont (NABEL, 2016, 9, 66–68) 30 km to the East of La Brévine using the immersion freezing method described by Conen et al. (2012). The $PM_{10}$-filter samples cover a time interval of 24 hours, from midnight to midnight. For the comparison of IP with INP we used the filter of the day when an IP measurement had started.

# 3 Results

The deposition rate of IP was estimated from the number of IP $n(IP)$ counted during the total exposure time $t$ (40 minutes for 60 minutes observation time) and the size of the area $A$ on the (imaginary) horizontal plane skimming the upper edge of the photographed volume, through which the detected IP may have passed (Figure 2). This area $A$ has the shape of a parallelogram. One side of this parallelogram ($b$) has the same length as the photographed area. The other side ($a$) is calculated from the height of the photographed area $f$ divided by the settling velocity of the IP ($v_{IP}$) and multiplied by the horizontal wind velocity ($v_{Wind}$). The angles of the parallelogram ($\alpha$) are defined by the angle in which the wind cuts the light curtain on the horizontal plane. The equation for the deposition rate is:

$$deposition\ rate\ D_{IP} = \frac{n(IP)}{A \cdot t} \quad , \quad (1.1)$$

where

$$A = a \cdot h = \left(\frac{f \cdot v_{wind} \cdot h}{v_{IP}}\right) = \left(\frac{f \cdot v_{wind} \cdot \sin(\alpha) \cdot b}{v_{IP}}\right) \quad . \quad (1.2)$$

With $h$ being the height of the parallelogram. We assume the settling velocity of IP to be 0.1 m s$^{-1}$ (Yagi, 1970). Wind velocity and prevailing wind direction have been determined from measurements of the meteorological station La Brévine (46.98° N, 6.62° O, 1050 m a.s.l.).

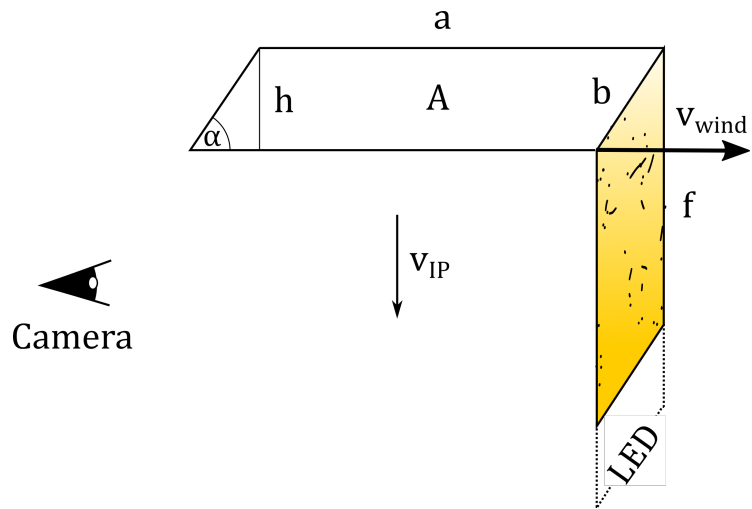

*Figure 2: Sketch illustrating the setup for detecting IP and the parameters used in the calculation of the IP deposition rate (Eq. 1.1 and 1.2). The upwards-facing LED unit creates the light curtain (yellow area, about 30 mm thick), in which ice particles are illuminated and scatter part of the light into the camera. The area A is the horizontal catchment area of IP whose traces were recorded by the camera.*

On 12.12.2015 the temperatures varied between -5.8 °C and -6.6 °C. In this temperature range (-5 °C to -7 °C) 1 INP m$^{-3}$ has been found on the PM$_{10}$-filter. The mean deposition rate of IP obtained with equation 1.1 was 0.08 m$^{-2}$ s$^{-1}$. At this deposition rate, the reservoir of INP in the lower 10 m of the fog layer would be depleted after two minutes (1 INP m$^{-3}$ * 10 m / 0.08 m$^{-2}$ s$^{-1}$). However, we observed on-going deposition of IP during one hour. Hence, the observed number of IP was an order of magnitude larger than that of INP in the same volume of air. Table 1 shows that this phenomenon was less pronounced on 09.03.2016 but still clearly detectable. Although the actual thickness of the fog layer may be greater than the presumed 10 m, temperature increases quickly with height (Vitasse et al., 2016). Assuming the coldest temperatures measured at 1 m above ground to extend to 10 m above ground is a conservative assumption in our context. In an inversion the temperature at 10 m above ground is higher than the temperature at 1 m above ground. The discrepancy between IP and INP would even be larger if the actual temperature profile was available and taken into account, because the number of INP activated decreases exponentially with increasing temperature (Petters & Wright, 2015).

*Table 1: Deposition rates of IP and conditions during the two measurement periods in radiation fog. For comparison we estimated INP deposition rates from INP number concentrations determined on $PM_{10}$-filters.*

|  | 12.12.2015 | 09.03.2016 |
|---|---|---|
| Measurement period (local time, UTC+1) | 22:56-23:56 | 23:07-00:35 |
| Exposure time $t$ [min] | 40 | 58 |
| Air temperature range $\Delta T$ [°C] | -5.8 to -6.6 | -3.1 to -9.2 |
| $v_{Wind}$ [m s$^{-1}$] | 1.3 | 0.66 |
| $\alpha$ | 45° | 45° |
| Presumed fog layer thickness $h_{fog}$ [m] | 10 | 10 |
| Number of IP counted during exposure time $n(IP)$ | 150 | 43 |
| Measured deposition rate of IP $D_{IP}$ (see Eq. 1.1) [IP m$^{-2}$ s$^{-1}$] | 0.08 | 0.03 |
| Measured INP on $PM_{10}$-filter activated in $\Delta T$ ($C_{INP}$) [m$^{-3}$] | 1 | 2 |
| Maximum possible deposition rate of INP $D_{INP} = \frac{C_{INP} \cdot h_{fog}}{t}$ [INP m$^{-2}$ s$^{-1}$] | 0.004 | 0.006 |
| INP-reservoir in $h_{fog}$ depleted after [min] | 2 | 10 |
| Ratio $\frac{D_{IP}}{D_{INP}}$ | 19 | 5.4 |

# 4 Discussion

During observations in supercooled radiation fog (-3.1 °C to -9.2 °C) we could quantify the deposition rate of ice particles with a relatively inexpensive setup. Deposition rates of IP were 19 (12.12.2015) and 5.4 (09.03.2016) times larger than expected from the reservoir of INP in the layer of radiation fog. The same phenomenon, several orders of magnitude more IP than INP, has previously been reported from observations in mixed-phase clouds at higher altitudes (Pruppacher & Klett, 1997; Wallace & Hobbs, 2006), and has recently been observed by Westbrook & Illingworth (2013), Lloyd et al. (2015), and Farrington et al. (2016).

There are several processes that might explain the discrepancy we observed in radiation fog:

*Underestimation of INP due to deactivation on filters*

An instability of INP on $PM_{10}$-filters could have influenced our results, leading to an underestimation of INP number concentrations in the air. While active bacterial INP are easily deactivated by chemical, anaerobic and heat stresses, INP from decayed leaf litter are stable over a wide range of time and stresses and some samples kept at room temperature showed the same ice nucleation activity after 30 years (Schnell, 2009). Deactivation can probably explain only a minor part of the observed difference.

*Underestimation of INP due to detection limits*

Mason (1971) states that most INP active at > -12 °C are too small to be detected by conventional sampling techniques but are still able to initiate IP formation. Pruppacher & Klett (1997) reckon that the discrepancy between INP- and IP-concentrations could be substantially lowered or even eliminated with adequate instrumentation. However, the quartz-fibre filters used in our study are very efficient in sampling even nanoparticles (Tsai et al., 2012). Hence, underestimation is an unlikely explanation of our observation.

*Ice multiplication*

As the Hallet-Mossop process prevails only with graupel particles (Field et al., 2016) it can be ruled out as a possible ice multiplication process during our observations in a shallow layer of radiation fog. However, the shattering of freezing water droplets, as described by Pander (2015) and Field et al. (2016), is not dependent on the occurrence of graupel particles (Pander, 2015). At -6 °C Takahashi & Yamashita (1970) observed a fragmentation frequency of 17%. This would imply 19-fold and 10-fold cascades of shattering droplets to explain the observations made on 12.12.2015 and on 09.03.2016, respectively. Hence, the fragmentation of freezing water droplets is unlikely to explain all, but may have contributed partly to the discrepancy between IP and INP we observed.

*Other processes*

The influence of hoar frost and blowing snow, which seems to have been important during observations on the Jungfraujoch (Farrington et al., 2016; Lloyd et al., 2015), can be ruled out during our observations due to the very strong inversion and the prevailing low wind velocities during observation.

If the temperature of INP was much lower than that of the surrounding air because of radiative cooling of the suspended particles, activated INP number concentrations might have been underestimated by assuming INP activation at air temperature (Mukund et al., 2014).

As INP are not suspended in the air but settling, they can enter the radiative fog layer from above, even though during strong inversions the gas exchange between the very stable nocturnal boundary layer and the air above is extremely small (Xia et al., 2011). During an observation period of 60 min, a $PM_{10}$-particle with a settling velocity of 0.005 m s$^{-1}$ (Heldstab et al., 2003) would sink 18 m. Assuming that all INP had an aerodynamic diameter of 10 µm ($PM_{10}$), the reservoir of INP producing IP in the radiation fog layer might therefore exceed the fog layer by 18 m. If turbulent diffusion above the fog layer is taken into account, this value could be substantially larger and might be large enough to explain much of the observed difference between IP and INP in our observations. It would also be in accordance with the larger difference found on the day with the greater average wind speed (12.12.2015).

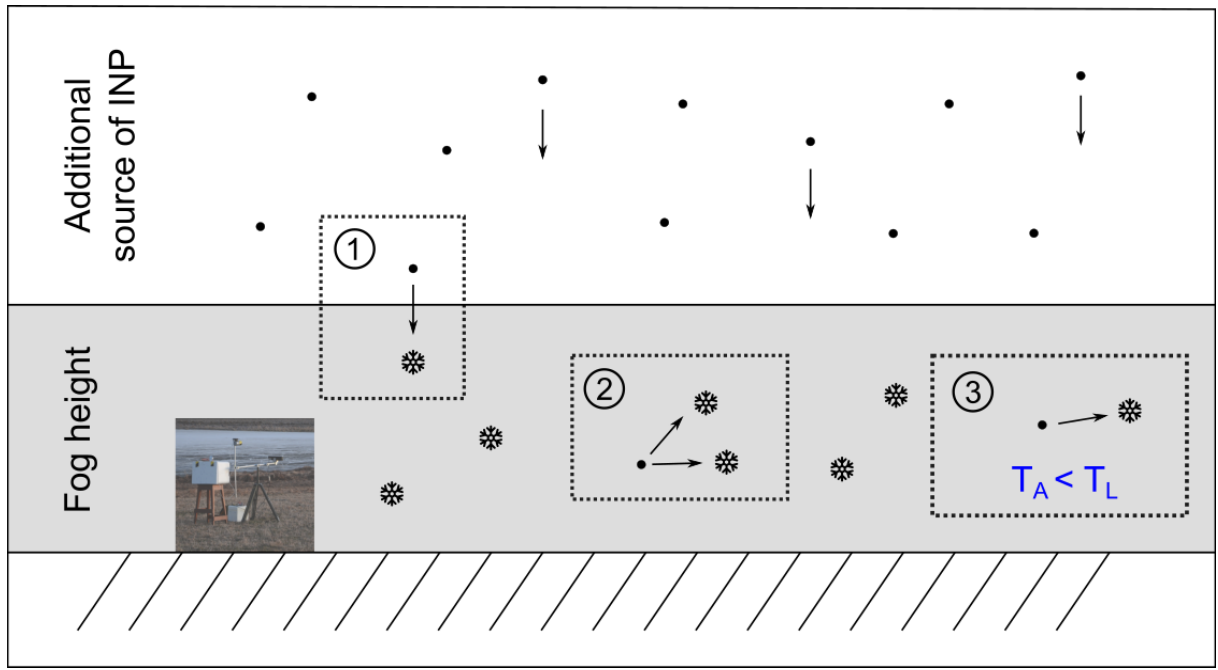

*Figure 3: Overview of the processes probably occurring during our observation of a much larger number of IP compared to INP. INP (•) are activated in the fog at sub-zero temperatures and IP (❄) form. There is an additional source of INP above the fog because the INP are settling. The maximum height of this source depends on the settling velocity of the particles and on the duration of the observation. The following processes are depicted: **(1)** Settling INP reaching the supercooled fog are activated and form IP. **(2)** Ice multiplication due to the fragmentation of freezing water droplets. **(3)** If aerosols are colder than the surrounding air, INP are activated earlier ($T_A$ = aerosol temperature; $T_L$ = air temperature). Photography: Lukas Zimmermann, Ice crystal drawing: Freepik, WWW.*

Several processes possibly taking place at the same time might explain the observed discrepancy between IP and INP deposition rates (Figure 3). It seems very unlikely that one process alone should be able to do so. We assume that a combination of the trend to underestimate INP concentrations measured on filters, the fragmentation of freezing water droplets, and INP deposited from above the fog may explain the observed discrepancy between IP and INP. Aerosols being colder than the surrounding air due to radiative cooling (Mukund et al., 2014) might have some influence as well.

## Conclusion

There are large uncertainties regarding the relation between INP and IP in clouds. With an unusual approach we managed to quantify IP deposition rates in a temperature range where measured data is rare (>-10 °C). Radiation fog constitutes an easily accessible form of a supercooled cloud, in which observations can be made for long enough to quantify the deposition rate of rare IP produced in a natural environment. The corresponding strong inversion provides a system that is relatively closed compared to other low or mid-level clouds. Still, entrainment from above can not be neglected.

## Acknowledgements

We thank NABEL for the provision of the PM$_{10}$-filters, Yann Vitasse and Geoffrey Klein for helpful discussions, Lukas Zimmermann for technical support and Alexandra Staubli-Buerge for help with the on-site measurements.